\documentclass[10pt,aps,pra,twocolumn,nofootinbib,superscriptaddress,floatfix]{revtex4-2}
\usepackage{amssymb,amsmath}
\usepackage{color}
\usepackage{hyperref}
\usepackage{listings}
\usepackage{physics}
\usepackage{amsmath}
\usepackage{graphicx}
\usepackage{caption}
\usepackage{subcaption}
\usepackage{bbold}
\usepackage{bbm}
\usepackage{booktabs}
\usepackage{makecell}
\usepackage{qcircuit}
\usepackage{multirow}
\usepackage{bm}
\usepackage[normalem]{ulem}
\usepackage{placeins}

\usepackage[toc,page]{appendix}

\usepackage[table,xcdraw,dvipsnames]{xcolor}

\hypersetup{
    unicode=false,          
    pdftoolbar=true,        
    pdfmenubar=true,        
    pdffitwindow=false,     
    pdfstartview={FitH},    
    pdfauthor={},     
    colorlinks=true,       
    linkcolor=blue,          
    citecolor=red,        
}

\definecolor{dkgreen}{rgb}{0,0.6,0}
\definecolor{gray}{rgb}{0.5,0.5,0.5}
\definecolor{mauve}{rgb}{0.58,0,0.82}

\newcommand{\UNITN}{{Dipartimento di Fisica, University of Trento,
    via Sommarive 14, I--38123, Povo, Trento, Italy}}
\newcommand{\TIFPA}{INFN-TIFPA Trento Institute of Fundamental Physics
    and Applications, Trento, Italy}

\begin{document}
\title{Lie Algebra-Based Quantum Optimal Controls Interpolation}

\author{Piero Luchi}
\email{piero.luchi@unitn.it}
\affiliation{\UNITN}

\author{Francesco Pederiva}
\affiliation{\UNITN}
\affiliation{\TIFPA}

\begin{abstract}
We present a framework combining Lie group theory and feed-forward neural
networks to efficiently generate quantum optimal control pulses for arbitrary
unitary operations in superconducting qubit systems, bypassing the need for
explicit optimization at inference time.
The exponential scaling of the Hilbert space dimension with qubit number makes
standard optimization approaches computationally prohibitive when large ensembles
of distinct propagators must be processed, a bottleneck that is particularly
acute in Trotterized quantum simulation.
Our method addresses this limitation by pre-computing a representative set of
control pulses via Lie group theory and training neural networks to map target
propagators to their corresponding pulses.
We demonstrate the approach on superconducting qubit systems of 2, 3, and 4
qubits, finding high reconstruction fidelity for specific combinations of Lie
algebra parameters.
As a physically motivated benchmark, we apply the methodology to reconstruct
control pulses for the Trotter propagators of a neutrino system undergoing
collective flavor oscillations.
The successful generalization across system types demonstrates that a single
model -- trained once on hardware-specific random propagators -- can serve as a
universal control-pulse generator for any target quantum system of compatible
Hilbert space dimension, offering a promising route toward scalable quantum
simulation.
\end{abstract}

\maketitle

\section{Introduction}\label{sec:intro}
Implementing a generic unitary operation in a quantum computer usually relies on
its decomposition into a sequence of one- and two-qubit unitary basis
operations~\cite{barenco1995gates}.
This approach is based on the well-known Solovay--Kitaev theorem
\cite{harrow2001solvay_kitaev_theorem}, stating that the number of elements into
which a given unitary operation can be approximated by decomposing it into a set of
elementary operations scales reasonably with its size.
Despite this property, quantum circuits resulting from such decomposition are often
too deep to be run within the typical decoherence time of present-day qubits. This results in the accumulation of errors that
individual basic operations introduce in the whole circuit, eventually spoiling
the result.

Quantum optimal control (QOC) techniques~\cite{glaser2015QOC,werschnik2007QOC}
were applied to this field as a possible route to reduce this error accumulation.
QOC helps to steer the time evolution of a quantum system (i.e.\ the qubits) to
realize the desired unitary operation.
This is obtained by optimizing a set of external control pulses, coupled to the
quantum system via control operators, such that they drive the system according to
the desired unitary.
This allows one to implement arbitrary unitary operations in a minimal number of
applications, reducing the duration of transformation and therefore the overall
accumulation of error. Two caveats are in order. First, QOC presupposes accurate
knowledge of the system Hamiltonian, an assumption that is non-trivial in practice.
Second, the flexibility of the approach comes at a computational cost that scales
exponentially with the number of qubits, i.e. with the Hilbert-space dimension,
a limitation that QOC shares with the gate-decomposition approach.

The cost of QOC becomes particularly limiting in the context of digital quantum
simulation. The Lloyd construction~\cite{lloyd1996universal}, together with its
higher-order Trotter--Suzuki refinements~\cite{childs2021trotter}, expresses the
time evolution generated by a many-body Hamiltonian as a product of short-time
propagators, one for each Trotter step.
A single simulation run therefore requires the realization of a large number of
\emph{distinct} unitaries, and exploring different evolution times, Hamiltonian
parameters, or physical observables multiplies this count further.
Re-running a full optimal-control routine independently for every target is
impractical, and motivates the search for methods that, after the optimization of
a representative set of pulses performed once and for all, can produce the
control sequence for any new target unitary at negligible additional cost.

A possible way to tackle this bottleneck is to combine QOC with machine
learning. Such a combination is one of the main lines of the broader effort to
apply artificial intelligence across the quantum-computing
stack~\cite{alexeev2025AI_QC}.
Reinforcement learning has been shown to discover control protocols
competitive with, or in some regimes superior to, gradient-based optimizers for
both state preparation~\cite{bukov2018RL_quantum_control} and gate
synthesis~\cite{niu2019deepRL_QOC}, while gradient-based and chopped-basis
schemes such as GRAPE~\cite{khaneja2005GRAPE,rowland2012GRAPE} and
CRAB~\cite{caneva2011CRAB} remain the standard tools for the underlying
single-target optimization.
More recent work has begun to address the multi-target setting directly:
Sauvage and Mintert~\cite{sauvage2022families} optimize a single parametrized
pulse family, represented by a neural network, across a continuous set of
target gates, Chadwick and Chong~\cite{chadwick2023QOC} re-optimize a set of
reference pulses so that linear interpolation between them yields
high-fidelity controls across the whole family, and pulse-level perspectives
on variational quantum algorithms~\cite{magann2021pulses_circuits} highlight
the value of recycling control resources across related tasks.
Neural networks have also been trained on optimal-control solutions to return
the controls of a new target gate directly: physics-informed networks
interpolate continuously parametrized one- and two-qubit gate
families~\cite{bhattacharyya2024NN_interpolation}, and a network trained on
GRAPE pulses generates the controls of arbitrary single-qubit gates on an NMR
processor~\cite{khaleghian2026NN_GRAPE}. In both cases the target family is
low-dimensional, with at most three parameters.
What is still largely missing is a framework that (i) exploits the
group-theoretic structure of the target unitaries to define a representative
training set, and (ii) learns the map from target propagator to control
pulses by regression, so that the trained model serves as a general-purpose
pulse generator across physically distinct simulation tasks.

In this work we propose such a framework, which we call Lie Algebra-based Control
Pulse Reconstruction (LA-CPR).
The starting observation is that every $N_q$-qubit propagator belongs to the
compact Lie group $SU(N)$, which can be sampled systematically through its
Lie algebra $\mathfrak{su}(N)$ expressed in the Pauli tensor basis.
We restrict the sampling to a tractable subset of generators corresponding to a
linear chain topology with Ising $ZZ$ couplings -- understood here as a
restriction on the unitaries to be synthesized rather than as a constraint on the
underlying hardware -- and draw random elements from a neighborhood of the group
identity.
For each sampled propagator a control pulse is optimized once via
GRAPE~\cite{khaneja2005GRAPE,rowland2012GRAPE}, building a dataset of
propagator--pulse pairs.
Two feed-forward neural networks are then trained to regress the in-phase and
quadrature components of the control envelope from the matrix entries of the
target unitary, so that at inference time the controls for any new target are
obtained by a single network evaluation, with no further optimization required.
We validate the method on superconducting qubit systems of $N_q = 1$--$4$ qubits
and, as a physically motivated benchmark, we apply it to the Trotter propagators
of neutrino collective flavor
oscillations~\cite{Hall2021neutrinos,amitrano2023neutrinos}, showing that a
network trained on hardware-generic samples generalizes to a quantum simulation
task it has never seen during training, building on and extending our earlier
work on pulse reconstruction~\cite{luchi2023CPR,turro2023coprocessing}.

\section{Methods}\label{sec:methods}

\subsection{Quantum Optimal Control}\label{subsec:quantum_optimal_control}
The general Hamiltonian of a qubit system is:
\begin{eqnarray}
    \label{eq:QC_general_hamiltonian}
    H_{QC} = H_0 + \sum_{k=1}^{N_{ctrl}}\epsilon_k(t)H_k.
\end{eqnarray}
$H_0$ is the drift term of the Hamiltonian, and it depends on the physical
properties of the device, such as the type of qubits (trapped ions, superconducting,
etc.), their couplings and connectivity.
$H_k$ are the drive Hamiltonians that describe how the external environment can
interact with the system.
The modulation in time of the strength of these drive Hamiltonians is described
by the $N_{ctrl}$ functions $\epsilon_k(t)$.
The number $N_{ctrl}$ of such controls depends on the specific hardware and the
connectivity between the qubits.

The time-evolution of the system during a time step $t_j$ of the pulses is given
by the propagator:

\begin{eqnarray}
    U_{t_j}=\exp{-i \left ( H_0 + \sum_{k=1}^{N_{ctrl}}\epsilon_k(t_j)H_k \right ) \Delta t },
\end{eqnarray}
hence the system's state $\ket{\psi(t)}$ at the final time $t_N$ is:

\begin{eqnarray}
    \ket{\psi(t_N)}=U_{t_N} U_{t_{N-1}}...U_{t_2}U_{t_1}\ket{\psi(t_0)}.
\end{eqnarray}
The complete propagator is therefore:
\begin{eqnarray}
    \label{eq:QC_complete_propagator}
    U&=&U_{t_N} U_{t_{N-1}}...U_{t_2}U_{t_1} \nonumber\\
    &=&\mathcal{T}\exp{-i \left ( H_0 + \sum_{j=0}^{N} \sum_{k=1}^{N_{ctrl}}\epsilon_k(t_j)H_k \right ) \Delta t },
\end{eqnarray}
where $\mathcal{T}\exp{}$ stands for time-ordered exponential.
\noindent The optimal control problem consists in finding the best control shapes
$\epsilon_k(t)$ such that, when applied to the quantum device for a total time
$t_N$, drive the system from the state $\ket{\psi(0)}$ to the desired state
$\ket{\psi(t_N)}$ within an acceptable error.

Among the many optimization procedures available, in this work we use the
GRadient Ascent Pulse Engineering (GRAPE) procedure
\cite{khaneja2005GRAPE,rowland2012GRAPE}.
In particular, given a target operator $U_{targ}$ (that drives the quantum system
state in the desired way), the optimal set of controls is found by adjusting
their single time-steps $\epsilon_k(t_j)$ $\forall k \in [0,N_{ctrl}],j\in [0,N]$
while maximizing the cost function called \emph{fidelity} $\mathcal{F}$:
\begin{eqnarray}\label{eq:fidelity}
\mathcal{F}=1-\tilde{\mathcal{D}},
\end{eqnarray}
where $\tilde{\mathcal{D}}$ is the normalized Hilbert-Schmidt norm,
\begin{eqnarray}
\tilde{\mathcal{D}}=\frac{1}{2} - \frac{1}{2d }\ \mathrm{Re} \ \mathrm{Tr}(U_{targ}^\dagger U),
\end{eqnarray}
with $U$ the complete propagator of Eq.~\eqref{eq:QC_complete_propagator} and
$d=\dim(U_{targ})$ the dimension of the matrices.
This metric quantifies the similarity of the two matrices by giving a value close
to one when they are similar and close to zero when they are different.
We thus need to solve the general optimization problem:

\begin{equation}\label{eq:general_opt_problem}
    \{\epsilon_k^*(t_j)\} = \underset{\{\epsilon_k(t_j)\}}{\arg\max} \;\mathcal{F}\!\left(U_{\mathrm{targ}},\, U(\{\epsilon_k(t_j)\})\right).
\end{equation}

The optimal control framework outlined above is general and applies to any form
of the qubit Hamiltonian $H_0$.
In the next subsection (Sec.~\ref{subsec:qubits_system_XY}) we specialize it to
the hardware considered in this work: a system of $N_q$ superconducting qubits
coupled to a common cavity bus resonator, operated in the dispersive regime and
driven by a global microwave control
field~\cite{blais2021cQED,koch2007transmon,blais2004cavity,wallraff2004strong}.

\subsection{Qubits System}\label{subsec:qubits_system_XY}

We now specialize the general control Hamiltonian of
Eq.~\eqref{eq:QC_general_hamiltonian} to the hardware considered in this work,
by deriving the effective qubit Hamiltonian below.

The starting point is the generalized Tavis-Cummings
model~\cite{tavis1968} within the rotating-wave approximation
(RWA)~\cite{blais2004cavity}.
By assuming that the detuning between the qubits and the cavity mode is much
larger than their respective coupling strengths, i.e., $|\omega_k - \omega_c|
\gg g_k$, we can perturbatively eliminate the cavity degrees of freedom using a
Schrieffer-Wolff transformation~\cite{blais2004cavity}.
In this dispersive regime, direct energy exchange between the qubits and the
cavity is suppressed.
Instead, the cavity mediates virtual photon exchanges, which manifest as an
always-on effective transverse interaction ($XY$-type exchange) between the
qubits~\cite{majer2007cavitybus,blais2007QED}.

To describe the driven dynamics, we move into a frame rotating at the angular
frequency of the global microwave drive, $\omega_d$.
In this rotating frame, the fast oscillating terms of the control field are
eliminated, leaving only the slowly varying pulse
envelopes~\cite{krantz2019superconducting_qubits}.
The drift, interaction, and control terms of
Eq.~\eqref{eq:QC_general_hamiltonian} take then the explicit form:
\begin{equation}
    \hat{H}_{QC}(t) = H_{\text{drift}} + H_{\text{int}} + H_{\text{ctrl}}(t),
    \label{eq:H_qubit_tot}
\end{equation}
where the drift, interaction, and control Hamiltonians are given by:
\begin{align}
    H_{\text{drift}} &= \sum_{k=1}^{N_q} \frac{\Delta_k}{2} \sigma_z^{(k)}, \label{eq:H_drift} \\
    H_{\text{int}} &= \sum_{\langle j,k \rangle} J_{jk} \left( \sigma_+^{(j)} \sigma_-^{(k)} + \sigma_-^{(j)} \sigma_+^{(k)} \right) \nonumber \\
    &= \sum_{\langle j,k \rangle} \frac{J_{jk}}{2} \left( \sigma_x^{(j)} \sigma_x^{(k)} + \sigma_y^{(j)} \sigma_y^{(k)} \right), \label{eq:H_int} \\
    H_{\text{ctrl}}(t) &= \frac{1}{2} \Omega_x(t) \sum_{k=1}^{N_q} \lambda_k \sigma_x^{(k)} \nonumber \\
    &\quad + \frac{1}{2} \Omega_y(t) \sum_{k=1}^{N_q} \lambda_k \sigma_y^{(k)}. \label{eq:H_ctrl}
\end{align}
Here, $N_q$ is the number of qubits.
The operators $\sigma_{x,y,z}^{(k)}$ are the standard Pauli matrices acting on
the $k$-th qubit, with raising and lowering operators defined as
$\sigma_+^{(k)} = |1\rangle\langle 0|$ and $\sigma_-^{(k)} = |0\rangle\langle 1|$.
Throughout this work we adopt natural units with $\hbar = 1$.

The drift Hamiltonian $H_\text{drift}$ describes the free precession of each
qubit in the rotating frame, where $\Delta_k = \omega_k - \omega_d$ is the
detuning between the bare transition frequency $\omega_k$ of the $k$-th qubit
and the global drive frequency $\omega_d$.
In this work, the drive frequency $\omega_d$ is chosen to be the average qubit
frequency, $\omega_d = \frac{1}{N_q}\sum_{k=1}^{N_q} \omega_k$, to minimize
the maximum detuning across the register.

The interaction Hamiltonian $H_\text{int}$ captures the exchange-type coupling
between pairs of qubits, mediated by dispersive interaction with the cavity.
The coupling rate $J_{jk}$ quantifies the effective flip-flop interaction
strength between qubits $j$ and $k$.
The sum runs over nearest-neighbor pairs in the connectivity graph of the device:
the angle-bracket notation $\langle j,k \rangle$ denotes an unordered pair of
neighboring qubits, so each interacting pair is counted exactly once.
In this work, we assume a chain topology for the qubit connectivity.

The control Hamiltonian $H_\text{ctrl}(t)$ represents the action of the global
microwave drive on the qubit register.
The time-dependent envelopes $\Omega_x(t)$ and $\Omega_y(t)$ are the slowly
varying Rabi frequencies corresponding to the in-phase (I) and quadrature (Q)
components of the pulse, respectively.
The dimensionless coefficient $\lambda_k$ accounts for the relative coupling
strength of the global drive field to each individual qubit, which may differ
due to variations in the physical layout of the device.
For simplicity, we set $\lambda_k = 1$ for all $k$ throughout this work.


\subsection{Lie Groups and Algebras}\label{subsec:lie_theory}

A \emph{Lie group} $G$ is a finite-dimensional smooth manifold endowed with a
smooth group structure. Its associated \emph{Lie algebra} $\mathfrak{g}$ is the
space of infinitesimal generators of the group, connected to $G$ through the
\emph{exponential map} $\exp:\mathfrak{g}\to G$, which allows any group element
to be written as $U(\gamma)=\exp\{-i\gamma\cdot T\}$, where the Hermitian
generators $T_p=i\,\partial U/\partial\gamma_p|_{\gamma\to 0}$ are such that
$\{-iT_p\}$ spans $\mathfrak{g}$~\cite{das2014lie}.

In the context of an $N_q$-qubit system, the relevant group is the special
unitary group $SU(N)$, where $N \equiv 2^{N_q}$ denotes the dimension of the
Hilbert space. $SU(N)$ is the group of $N \times N$ unitary matrices with
determinant $1$; it is a real Lie group of dimension $N^2-1$, whose Lie
algebra $\mathfrak{su}(N)$ consists of traceless anti-Hermitian $N\times N$
matrices. Any unitary transformation $U$ acting on $N_q$ qubits belongs to
$SU(N)$ and can be written as the exponential of its generator in
$\mathfrak{su}(N)$~\cite{nielsen2000,das2014lie}:
\begin{eqnarray}
    U &=& \exp\!\left(-i H\right) \in SU(N),\\
    H &=& \sum_{k=1}^{N^2 - 1} \gamma_k\, T_k,
    \qquad -i\,H \;\in\; \mathfrak{su}(N),
\end{eqnarray}
where the Hermitian $\{T_k\}_{k=1}^{N^2-1}$ are such that $\{-iT_k\}$ is an
orthonormal basis of $\mathfrak{su}(N)$ with respect to the Hilbert--Schmidt
inner product, satisfying~\cite{chew2020,shah2022}
\begin{equation}
    \operatorname{Tr}\!\left(T_j^\dagger T_k\right) = N\,\delta_{jk}.
\end{equation}
The parameters $\gamma_k$ are \emph{free real numbers}. Taken one at a time,
each of them is $2\pi$-periodic,
\begin{equation}
    e^{-i(\gamma_k + 2\pi)T_k} = e^{-i\gamma_k T_k},
\end{equation}
as follows from $T_k^2 = \mathbbm{I}$ and the identity
$e^{-i\gamma T} = \cos(\gamma)\,\mathbbm{I} - i\sin(\gamma)\,T$~\cite{nielsen2000}.
Restricting every coefficient to the interval $\gamma_k \in [-\pi, +\pi]$,
symmetric about the origin, is sufficient to encompass the entire group, and
this box is adopted as the sampling domain throughout. Such a box actually
over-covers $SU(N)$. In fact, the exponential map
$\exp\colon \mathfrak{su}(N) \to SU(N)$ is surjective~\cite{brocker1985}, being
$SU(N)$ compact and connected, but not injective, so that distinct coefficient
vectors can define the same group element. For simplicity we accept this
redundancy, which allows each coefficient to be drawn independently and
uniformly from a single interval, as done in the sampling procedure of the
subsequent sections of this work.

\subsubsection{Pauli Tensor Basis Decomposition}

The generators $T_k$ are all tensor products of Pauli matrices
$\{\mathbbm{I}, \sigma_x, \sigma_y, \sigma_z\}^{\otimes N_q}$ excluding $\mathbbm{I}^{\otimes N_q}$,
and are naturally classified by their \emph{locality} $\ell$, i.e., the
number of qubits on which they act non-trivially~\cite{chew2020,shah2022}.
The Hamiltonian takes the explicit form
\begin{equation}
\label{eq:H_general}
    H = \sum_{\ell=1}^{N_q}
        \sum_{i_1 < \cdots < i_\ell}
        \sum_{a_1,\ldots,a_\ell}
        \gamma_{a_1 \cdots a_\ell}^{(i_1 \cdots i_\ell)}\,
        \sigma_{a_1}^{(i_1)} \otimes \cdots \otimes \sigma_{a_\ell}^{(i_\ell)},
\end{equation}
with $a_j \in \{x, y, z\}$. The number of $\ell$-qubits terms is
$\binom{N_q}{\ell} \cdot 3^\ell$, and summing over all locality levels
gives
\begin{equation}
    \sum_{\ell=1}^{N_q} \binom{N_q}{\ell} 3^\ell
    = (1+3)^{N_q} - 1
    = 4^{N_q} - 1
    = N^2 - 1,
\end{equation}
which equals the dimension of $\mathfrak{su}(N)$. As the $N^2 - 1$
parameters $\gamma_k$ range freely over $[-\pi, +\pi]$,
Eq.~\eqref{eq:H_general} covers the entire group $SU(N)$.

The $\ell$-qubits block captures genuinely $\ell$-partite correlations,
irreducible to lower-order terms. Discarding the
$N_q$-qubits terms, restricts $H$ to a subspace of dimension $N^2 - 1 - 3^{N_q}$,
which generates only a \emph{proper subgroup} of $SU(N)$.

\begin{table}[h]
\centering
\caption{Locality structure of the Pauli tensor basis for general $N_q$.}
\begin{tabular*}{0.95\columnwidth}{@{\extracolsep{\fill}}lll@{}}
\toprule
Locality $\ell$ & Operators & No.\ of terms \\
\midrule
1-qubits & $\sigma_a^{(i)}$ & $3N_q$ \\
2-qubits & $\sigma_a^{(i)}\otimes\sigma_b^{(j)}$, $i<j$ & $9\binom{N_q}{2}$ \\
$\vdots$ & $\vdots$ & $\vdots$ \\
$N_q$-qubits & $\sigma_{a_1}^{(1)}\otimes\cdots\otimes\sigma_{a_{N_q}}^{(N_q)}$ & $3^{N_q}$ \\
\midrule
\textbf{Total} & & $4^{N_q} - 1$ \\
\bottomrule
\end{tabular*}
\end{table}

\subsubsection*{\texorpdfstring{Example: $N_q=2$, $SU(4)$}{Example: Nq=2, SU(4)}}

For $N_q=2$, the algebra $\mathfrak{su}(4)$ has dimension
$4^2 - 1 = 15$~\cite{zhang2003,khaneja2001}.
The Hamiltonian expands as
\begin{equation}
    H = \underbrace{\sum_{i=1}^{2}\sum_a
        \gamma_a^{(i)} \sigma_a^{(i)}}_{\text{1-qubits: 6 terms}}
      + \underbrace{\sum_{a,b}
        \gamma_{ab}\, \sigma_a^{(1)}\otimes\sigma_b^{(2)}}_{\text{2-qubits: 9 terms}}.
\end{equation}

\begin{table}[h]
\centering
\caption{Pauli basis for $N_q=2$, $\dim\mathfrak{su}(4)=15$.}
\begin{tabular*}{0.95\columnwidth}{@{\extracolsep{\fill}}lll@{}}
\toprule
Locality & Operators & No.\ of terms \\
\midrule
1-qubits & $\sigma_a^{(1)},\,\sigma_a^{(2)}$ & 6 \\
2-qubits & $\sigma_a^{(1)}\otimes\sigma_b^{(2)}$ & 9 \\
\midrule
\textbf{Total} & & \textbf{15} \\
\bottomrule
\end{tabular*}
\end{table}

The analogous decomposition for $N_q=3$ ($\dim\mathfrak{su}(8)=63$),
which exhibits the additional 3-qubits locality block,
is reported in Appendix~\ref{App:pauli_basis_Nq3}.

\subsubsection{Simplifications in Qubits Chain Topology}\label{subsec:simplif_chain}
For sampling purposes, it is convenient to restrict the family of unitaries
that we ask LA-CPR to synthesize, rather than to draw arbitrary elements of
$SU(N)$. We impose two successive restrictions on this synthesized set,
each of which reduces the number of free parameters labeling the sample
while preserving universality. We stress that these restrictions act on
the \emph{target} unitaries used to build the training dataset: they do
\emph{not} constrain the underlying qubit Hamiltonian of
Eq.~\eqref{eq:H_qubit_tot}, which retains its full XY interaction and is
driven by global $\Omega_x(t),\Omega_y(t)$ envelopes throughout.

\paragraph{First simplification: linear chain topology.}
We first restrict the target unitaries to those generated by operators
acting on single qubits or on neighboring pairs of a linear chain
$1 \!-\! 2 \!-\! \cdots \!-\! N_q$. The corresponding generators span the
real vector space
\begin{equation}\label{eq:dynamic_algebra_g}
    \mathcal{V} = \mathrm{span}_{\mathbb{R}}\bigl\{\,
        i\sigma_a^{(i)},\quad
        i\sigma_a^{(i)}\otimes\sigma_b^{(i+1)}
    \bigr\},
\end{equation}
with $a,b \in \{x,y,z\}$. The generators acting on the pair $(i,i{+}1)$
span $\mathfrak{su}(4)_{i,i+1}$. Each site contributes three single-qubit
generators and each pair nine two-qubit ones, so that
\begin{equation}
    \dim\,\mathcal{V}
    = \underbrace{3N_q}_{\text{single-qubit}}
    + \underbrace{9(N_q-1)}_{\text{two-qubit}}
    = 12N_q - 9.
\end{equation}
For $N_q = 2$ this equals $\dim\,\mathfrak{su}(4) = 15$, and $\mathcal{V}$
is the whole algebra. For $N_q \geq 3$, instead, $12N_q - 9 < N^2 - 1$:
$\mathcal{V}$ is a proper subspace of $\mathfrak{su}(N)$, containing no
Pauli string that acts non-trivially on two non-adjacent qubits.

The subspace $\mathcal{V}$ is, however, not closed under the commutator.
Nested Lie brackets of nearest-neighbor generators produce effective
long-range couplings mediated by the intermediate qubits, and the Lie
algebra generated by $\mathcal{V}$, i.e.\ the dynamical Lie
algebra~\cite{zeier2011,dalessandro2021,wiersema2024DLA} of the chain, is the full
$\mathfrak{su}(N)$. Hence~\cite{jurdjevic1972,dalessandro2021}
\begin{equation}
    \mathrm{Lie}(\mathcal{V}) = \mathfrak{su}(N),
    \qquad
    \langle e^{\mathcal{V}} \rangle = SU(N),
\end{equation}
that is, every element of $SU(N)$ is a finite product of exponentials of
elements of $\mathcal{V}$. The chain topology thus reduces the number of
generators from $N^2 - 1$ to $12N_q - 9$ while preserving universality.
An explicit Pauli-string example of a long-range generator obtained in
this way is given in
Appendix~\ref{App:simpl_lie_alg}.

\paragraph{Second simplification: Ising $ZZ$ coupling.}
We further restrict the two-qubit generators to
$\sigma_z^{(i)}\otimes\sigma_z^{(i+1)}$ alone, retaining all the
single-qubit ones. The corresponding generators span
\begin{equation}
    \mathcal{V}_{\mathrm{ZZ}}
    = \mathrm{span}_{\mathbb{R}}\bigl\{\,
        i\sigma_a^{(i)},\quad
        i\sigma_z^{(i)}\otimes\sigma_z^{(i+1)}
    \bigr\} \subset \mathcal{V},
\end{equation}
with $a \in \{x,y,z\}$, and
\begin{equation}
    \dim\,\mathcal{V}_{\mathrm{ZZ}}
    = \underbrace{3N_q}_{\text{single-qubit}}
    + \underbrace{(N_q-1)}_{\text{two-qubit}}
    = 4N_q - 1,
\end{equation}
which is smaller than $N^2 - 1$ for every $N_q \geq 2$.

A single entangling generator per pair nevertheless suffices for
universality when combined with arbitrary single-qubit
rotations~\cite{khaneja2001,ramakrishna1995}: the eight two-qubit
directions $i\sigma_a^{(i)}\otimes\sigma_b^{(i+1)}$, $(a,b)\neq(z,z)$,
missing from $\mathcal{V}_{\mathrm{ZZ}}$ are recovered by commuting the
$ZZ$ generators with single-qubit ones. The Lie algebra generated by
$\mathcal{V}_{\mathrm{ZZ}}$ thus contains $\mathcal{V}$, and therefore
also the whole Lie algebra generated by $\mathcal{V}$, which is
$\mathfrak{su}(N)$:
\begin{equation}
    \mathrm{Lie}(\mathcal{V}_{\mathrm{ZZ}}) = \mathfrak{su}(N),
    \qquad
    \langle e^{\mathcal{V}_{\mathrm{ZZ}}} \rangle = SU(N).
\end{equation}
The $ZZ$ restriction thus reduces the number of generators from
$12N_q - 9$ to $4N_q - 1$ while preserving universality. An explicit
Pauli-string example showing how a $YZ$-type term is generated from $ZZ$
and single-qubit rotations is given
in Appendix~\ref{App:simpl_lie_alg}.

Based on both the chain-topology and the $ZZ$-only assumptions, the
elementary building block we adopt is therefore
\begin{eqnarray}
\label{eq:H_chain}
    H_{\text{chain}} =&&
    \underbrace{
        \sum_{i=1}^{N_q} \sum_{a \in \{X,Y,Z\}}
        \gamma_a^{(i)}\, \sigma_a^{(i)}
    }_{\text{1-qubit: } 3N_q \text{ terms}}   \nonumber\\
     &&+
    \underbrace{
        \sum_{i=1}^{N_q-1}
        \gamma_{ZZ}^{(i,i+1)}\,
        \sigma_z^{(i)} \otimes \sigma_z^{(i+1)}
    }_{\text{ZZ-only: } N_q-1 \text{ terms}},
\end{eqnarray}
involving $4N_q - 1$ independent real parameters
$\{\gamma_a^{(i)},\, \gamma_{ZZ}^{(i,i+1)}\} \subset [-\pi,+\pi]$.
By composing multiple layers of the form~\eqref{eq:H_chain}, any
target unitary $U_{\text{target}} \in SU(N)$ can be approximated
to arbitrary precision, at the cost of increased circuit
depth~\cite{dawson2006,brylinski2002}.

\subsection{Continuous Family of Controls}
Before turning to the LA-CPR method, we illustrate the property
that underpins it: the control pulses associated with neighboring points of
the Lie group depend smoothly on the algebra parameters.
We work this out explicitly for $SU(2)$, where the construction can be
carried through in closed form.

\begin{figure}[!htbp]
   \centering
  \includegraphics[width=\linewidth]{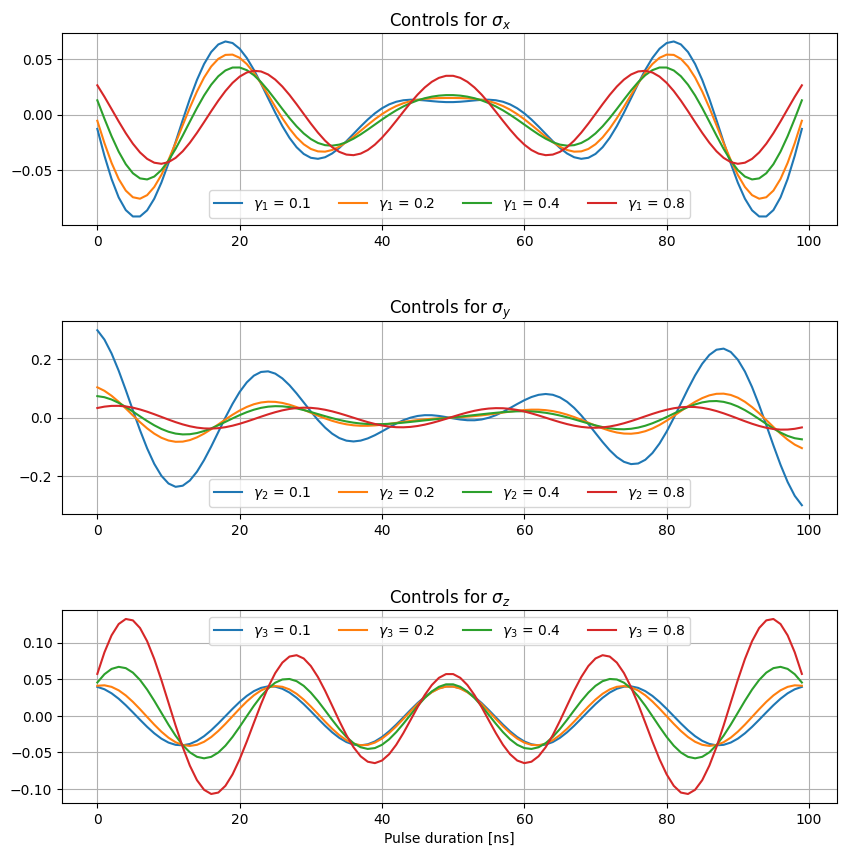}
   \caption{Example of controls $\epsilon_1(t)$ for different values of the parameters $\gamma_1,\gamma_2,\gamma_3$ in the generator of $SU(2)$.
   Small changes in the parameters lead to small changes in the controls, which is a consequence of the smooth manifold structure of $SU(2)$.
   In this work this feature is crucial to interpolate between controls.}
   \label{fig:example_controls}
\end{figure}

A generic element of $SU(2)$ can be written as the exponential of an
element of $\mathfrak{su}(2)$,
\begin{equation}
    U(\gamma_1,\gamma_2,\gamma_3)
      = \exp\bigl[i(\gamma_1\sigma_x+\gamma_2\sigma_y+\gamma_3\sigma_z)\bigr],
\end{equation}
with $\sigma_{x,y,z}$ the Pauli matrices. Using the single-generator
identities
\begin{align}
    e^{i\gamma_1\sigma_x}&=\begin{pmatrix} \cos\gamma_1 & i\sin\gamma_1  \\ i\sin\gamma_1 & \cos\gamma_1 \end{pmatrix}, \\
    e^{i\gamma_2\sigma_y}&=\begin{pmatrix} \cos\gamma_2 & \sin\gamma_2 \\ -\sin\gamma_2 & \cos\gamma_2 \end{pmatrix}, \\
    e^{i\gamma_3\sigma_z}&=\begin{pmatrix} e^{i\gamma_3}& 0  \\ 0 & e^{-i\gamma_3} \end{pmatrix},
\end{align}
and introducing the shorthand $c_j \equiv \cos\gamma_j$,
$s_j \equiv \sin\gamma_j$, the ordered product reads
\begin{eqnarray}
U &=& e^{i\gamma_1\sigma_x}\, e^{i\gamma_2\sigma_y}\, e^{i\gamma_3\sigma_z} \nonumber \\
&=& \begin{pmatrix}
e^{i\gamma_3}(c_1 c_2 - i s_1 s_2) & e^{-i\gamma_3}(c_1 s_2 + i s_1 c_2) \\[4pt]
e^{i\gamma_3}(i s_1 c_2 - c_1 s_2) & e^{-i\gamma_3}(c_1 c_2 + i s_1 s_2)
\end{pmatrix}.
\end{eqnarray}
The matrix entries are smooth functions of $(\gamma_1,\gamma_2,\gamma_3)$,
reflecting the smooth manifold structure of $SU(2)$. Consequently, the
controls $\epsilon_1(t)$ obtained from
Eq.~\eqref{eq:QC_complete_propagator} for any target $U(\gamma_1,\gamma_2,\gamma_3)$ inherit this
smooth dependence: nearby points in the algebra map to nearby pulse
shapes. Fig.~\ref{fig:example_controls} shows the GRAPE-optimized
controls obtained for several representative choices of
$(\gamma_1,\gamma_2,\gamma_3)$, illustrating how small variations of the
algebra coefficients translate into small, continuous variations of the
pulse envelope. It is this continuity that allows a feed-forward
network, trained on a discrete sample of $(U,\Omega(t))$ pairs, to
interpolate to unseen targets.

\subsection{Quantum Systems Simulation}

It is well established that the propagators $U$ of a quantum system belong to
the unitary group $U(N)$ or, upon neglecting a global phase---which leaves the
expectation values of all observables unchanged---to its subgroup, the special
unitary group $SU(N)$
\cite{georgi1982lie_groups1,hall2013lie_groups2}.
The most immediate example, corresponding to $N=2$, is provided by the Pauli
matrices introduced above, which describe the interaction of a particle's spin
with an external electromagnetic field and equally characterize the rotations of
a single qubit.
More generally, for a quantum system whose Hilbert space is $N$-dimensional,
the group $SU(N)$ describes the full set of admissible transformations of that
closed quantum system.

In a quantum computer of $N_q$ qubits, whose Hilbert space
$(\mathbb{C}^2)^{\otimes N_q}$ has dimension $N = 2^{N_q}$, the implementable
operations are the elements of $SU(N)$.
The problem of simulating the dynamics of a quantum system with propagator
$U_\mathrm{sys} \in SU(N)$ on a quantum computer therefore reduces to finding
an appropriate sequence of single- and multi-qubit operations, i.e.\ elements
of $SU(2^k)$ acting on subsets of $k$ qubits, such that their combined action
reproduces the effect of an
arbitrary target unitary $U_\mathrm{sys}$.
In the framework of quantum optimal control, these operations are realized as
optimized microwave pulse sequences that drive the qubit register into the
desired target state.

A standard approach to quantum simulation consists of decomposing the system
propagator as $U_\mathrm{sys} = e^{-i\,\delta t\, H_\mathrm{sys}}$ and
determining the corresponding control pulses via an optimization algorithm,
such as the one described in Sec.~\ref{subsec:quantum_optimal_control}.
A significant limitation of this approach emerges when the
Hamiltonian---and consequently the propagator---are time-dependent, i.e.,
$U_\mathrm{sys}(t) = e^{-i\,\delta t\, H_\mathrm{sys}(t)}$.
In this case, a new set of control pulses must be independently optimized for
each time step $U_\mathrm{sys}(t_i)$ of the simulation
\cite{luchi2023CPR,turro2023coprocessing}.
Furthermore, quantum simulations typically require computing the time evolution
from a large number of distinct initial configurations in order to fully
characterize the system, which further multiplies the number of required
optimization runs.
Taken together, these factors can rapidly offset the benefits of an optimal
control strategy.

\section{Lie Algebra Based Control Pulse Reconstruction}\label{sec:LA-CPR_method}

For a quantum device whose drift and control Hamiltonians close, under
iterated Lie brackets, on the full algebra $\mathfrak{su}(N)$, any unitary
$U \in SU(N)$ can in principle be realized by a suitable control sequence
$\epsilon(t)$~\cite{jurdjevic1972,ramakrishna1995,dalessandro2021}.
Such a sequence is accessible to standard optimization procedures
(e.g.\ GRAPE).
Conversely, the propagators of every closed quantum system whose Hilbert space
has dimension $N$ belong to $SU(N)$: the abstract group $SU(N)$ thus provides
a system-agnostic parametrization of all target unitaries of fixed dimension.

The method we propose exploits this correspondence.
Given a hardware register of $N_q$ qubits -- for which $N = 2^{N_q}$ -- we
sample a representative set of propagators $\{U_j\} \subset SU(N)$ via
Eq.~\eqref{eq:H_chain}, compute the corresponding GRAPE-optimized controls
once and for all, and train a feed-forward neural network to learn the map
$U \mapsto \epsilon(t)$.
Once trained, the network returns the controls for any new target
$\tilde{U} \in SU(N)$ at neural network inference cost,
including propagators originating
from physical systems other than the qubit register itself, provided their
Hilbert space dimension matches $N = 2^{N_q}$.

The sampling of the group elements builds on the Lie algebra framework
summarized in Sec.~\ref{subsec:lie_theory}: we use the parametrization
\begin{equation}
\label{eq:LA_sampling_U}
    U = \exp\!\left(-i H_{\text{chain}}\right),
\end{equation}
where $H_{\text{chain}}$ is the chain Hamiltonian of
Eq.~\eqref{eq:H_chain} and the coefficients
$\{\gamma_a^{(i)},\,\gamma_{ZZ}^{(i,i+1)}\}$ are the free real parameters
of the sampled unitaries.

The optimal-control problem of Eq.~\eqref{eq:general_opt_problem} is solved
here for the specific qubit Hamiltonian of Eq.~\eqref{eq:H_qubit_tot}, whose
control term $H_{\text{ctrl}}(t)$ is parametrized by the in-phase and
quadrature envelopes $\Omega_x(t),\,\Omega_y(t)$ of Eq.~\eqref{eq:H_ctrl}.
For this reason the controls reconstructed by LA-CPR are denoted
$\Omega_x(t),\,\Omega_y(t)$, rather than the generic $\epsilon_k(t)$ of
Sec.~\ref{subsec:quantum_optimal_control}.

The full \emph{Lie Algebra-based Control Pulse Reconstruction} (LA-CPR)
procedure proceeds in four steps:
\begin{enumerate}
    \item \textbf{Sampling.} Build a dataset of propagators $\{U_j\}$ from
    Eq.~\eqref{eq:LA_sampling_U}, drawing each coefficient
    $\gamma_a^{(i)},\,\gamma_{ZZ}^{(i,i+1)}$ independently from the uniform
    distribution on $[-z, z]$, with fixed $z \in \mathbb{R}$.
    \item \textbf{Optimization.} For each $U_j$, run GRAPE to obtain the
    in-phase and quadrature control envelopes $\Omega_x^j(t)$ and
    $\Omega_y^j(t)$, yielding the paired dataset
    $\{(U_j,\,\Omega_x^j,\,\Omega_y^j)\}$.
    \item \textbf{Training.} Train two feed-forward neural networks ---
    one for $\Omega_x(t)$, one for $\Omega_y(t)$ -- on the dataset, so
    that each network learns the map $U \mapsto \Omega_a(t)$ with $a \in [x,y]$.
    \item \textbf{Inference.} For any new target
    $\tilde{U} \notin \{U_j\}$, the trained networks return the
    reconstructed controls $\tilde{\Omega}_x(t),\,\tilde{\Omega}_y(t)$ in
    a single forward pass, with no further GRAPE optimization required.
\end{enumerate}

The GRAPE runs in step~2 are warm-started from the controls associated
with the propagator obtained by setting all coefficients of
$H_{\text{chain}}$ to $\gamma_a^{(i)},\gamma_{ZZ}^{(i,i+1)} = 0.1$.
This choice weights every generator equally and provides a balanced
reference unitary from which the rest of the dataset is reconstructed.

Implementation details of LA-CPR and the neural-network architecture are
deferred to Appendix~\ref{App:LA-CPR_details}.

\subsubsection{Datasets}

An appropriate dataset is essential for LA-CPR to have good performance.
As specified in Step 1 of the method, the Lie algebra coefficients
$\{\gamma_a^{(i)},\,\gamma_{ZZ}^{(i,i+1)}\}$ entering
Eq.~\eqref{eq:LA_sampling_U} are drawn independently from a uniform
distribution on the interval $[-z, z]$.
The choice of $z$ controls the region of the group manifold explored by the
dataset: a wider interval encompasses a larger and more diverse set of group
elements, while a narrower one restricts the sample to transformations closer
to the identity.

A wider interval, however, also makes the reconstruction task more
demanding for LA-CPR, as shown in the following.
For this reason, we restrict our analysis to three intervals,
corresponding to $z = \pi/8$, $\pi/4$, and $\pi/2$, which probe regions of
increasing extent around the identity.

Regarding the temporal structure of the control pulses, short pulses are
preferable as they keep the manipulation time well below the coherence times
of the qubits.
On the other hand, encoding all the information required for a faithful
transformation -- especially for larger qubit registers -- may require a
minimum pulse duration.
As a compromise between these two requirements, we fix the pulse duration to
$\tau = 150\,\mathrm{ns}$ for all datasets considered in this work.

\section{Results}\label{sec:results}

We evaluate LA-CPR on quantum systems of $N_q = 2, 3, 4$ qubits, each described
by Eq.~\eqref{eq:H_qubit_tot}, using the three datasets ($z = \pi/8,\,\pi/4,\,\pi/2$)
introduced in the previous section.
This setup lets us probe how the performance of the method depends both on the
system size and on the extent of the explored region of the group manifold.
The full list of numerical parameters used in the simulations, including the
qubit frequencies, the coupling strengths and the control sampling frequency,
is reported in Tab.~\ref{tab:system_numeric_param}.

\begin{table}[h]
\centering
\caption{Numerical parameters used in the simulations.}
\begin{tabular}{llll}
\hline
\textbf{Parameter} & \textbf{Symbol} & \textbf{Value} & \textbf{Unit} \\
\hline
\multicolumn{4}{l}{\textit{Time Parameters}} \\
Pulse duration       & $\tau$                & 150  & ns \\
Sampling frequency   & $f_s$                 & 2.0  & GHz \\
Number of time steps & $N_t = \tau\, f_s$    & 300  & -- \\
Time step            & $\delta t = \tau/N_t$ & 0.5  & ns \\
\hline
\multicolumn{4}{l}{\textit{Cavity Parameters}} \\
Cavity frequency    & $\omega_c$      & $2.5 \times 2\pi$ & GHz \\
Drive frequency     & $\omega_d$      & $\frac{1}{N_q} \sum_{i=0}^{N_q - 1} \omega_i$ & GHz\\
\hline
\multicolumn{4}{l}{\textit{Qubit Frequencies}} \\
Qubit 0 & $\omega_0$ & $1.0 \times 2\pi$ & GHz \\
Qubit 1 & $\omega_1$ & $1.1 \times 2\pi$ & GHz \\
Qubit 2 & $\omega_2$ & $1.2 \times 2\pi$ & GHz \\
Qubit 3 & $\omega_3$ & $1.3 \times 2\pi$ & GHz \\
\hline
\multicolumn{4}{l}{\textit{Coupling Strengths}} \\
Coupling qubits 0--1 & $J_{01}$ & $0.035 \times 2\pi$ & GHz \\
Coupling qubits 1--2 & $J_{12}$ & $0.040 \times 2\pi$ & GHz \\
Coupling qubits 2--3 & $J_{23}$ & $0.045 \times 2\pi$ & GHz \\
\hline
\end{tabular}
\label{tab:system_numeric_param}
\end{table}

\begin{figure*}[ht]
    \centering
    \includegraphics[width=\linewidth]{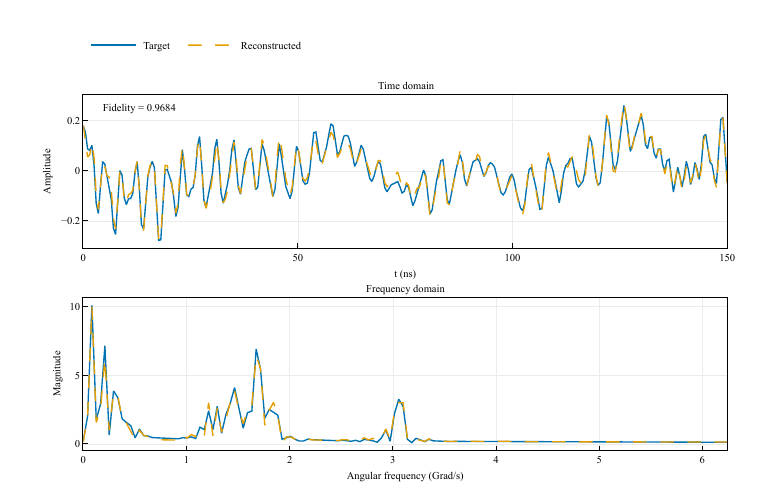}
    \caption{%
        Representative example of the control field $\Omega_x(t)$ reconstructed
        by the LA-CPR method for a 3-qubit system with $z = \pi/4$.
        \emph{Upper panel}: time-domain profile of the exact (target) control and
        its LA-CPR reconstruction; the inset reports the corresponding fidelity.
        \emph{Lower panel}: frequency spectra of both signals.
        The peak at ${\sim}0.03\ \mathrm{GHz}$ reflects the qubit--qubit
        interaction frequencies (see Tab.~\ref{tab:system_numeric_param});
        peaks above $1\ \mathrm{GHz}$ correspond to individual qubit transition
        frequencies; the feature near $3\ \mathrm{GHz}$ is a second-order
        harmonic artifact introduced by the optimizer due to the finite temporal
        discretization of the pulse.}
    \label{fig:example_control_and_spectrum}
\end{figure*}

The quality of the control pulses reconstructed by LA-CPR is assessed through
the \emph{fidelity} (Eq.~\eqref{eq:fidelity}) between the target and the
reconstructed propagators.
Given a target propagator $\tilde{U}$ from the test set, the trained neural networks of
the LA-CPR method return the control fields $\Omega_x(t),\,\Omega_y(t)$, from which the
reconstructed propagator $U_{\mathrm{recon}}$ is obtained as the discretized,
time-ordered evolution
\begin{equation}
\label{eq:QC_complete_propagator_specialized}
U_{\mathrm{recon}} = \mathcal{T} \exp \left[ -i \left( H_0 + \sum_{j=1}^{N_t} H_{\mathrm{ctrl}}(t_j) \right) \delta t \right],
\end{equation}
where $H_0 = H_{\mathrm{drift}} + H_{\mathrm{int}}$
(Eqs.~\eqref{eq:H_drift}--\eqref{eq:H_int}) is the time-independent part and
$H_{\mathrm{ctrl}}(t_j) = H_{\mathrm{ctrl}}\!\left(\Omega_x(t_j),\Omega_y(t_j)\right)$
is the control term of Eq.~\eqref{eq:H_ctrl} evaluated on the reconstructed
pulses at the $N_t$ discretization times $t_j$, with time step $\delta t$
(see Tab.~\ref{tab:system_numeric_param}).
The reconstruction quality is then quantified by the fidelity
$\mathcal{F}(\tilde{U}, U_{\mathrm{recon}})$.

\subsubsection{Graphical insight: control pulse and its spectrum}

Prior to the quantitative analysis of the LA-CPR method, we present a qualitative
examination of a representative control pulse.
Figure~\ref{fig:example_control_and_spectrum} shows an example of the control
field $\Omega_x(t)$ for a 3-qubit system with $z = \pi/4$.
The upper panel displays the exact control in the time domain alongside its
LA-CPR reconstruction, with the corresponding reconstruction fidelity indicated
in the inset.
The lower panel presents the frequency spectra of both the exact and reconstructed
controls.

Several features of the spectrum are noteworthy.
A peak at low frequencies (${\sim}0.03\ \mathrm{GHz}$) corresponds to the
qubit--qubit interaction frequencies listed in Tab.~\ref{tab:system_numeric_param},
which are around 35~MHz.
Additional peaks above $1\ \mathrm{GHz}$ are associated with the individual
qubit transition frequencies.
Finally, a peak near $3\ \mathrm{GHz}$ is identified as a harmonic artifact
introduced by the optimizer due to the finite temporal discretization of the
control pulse; its frequency is approximately twice the mean qubit frequency of
${\sim}1.5\ \mathrm{GHz}$, consistent with a second-order harmonic.

\subsubsection{Fidelity and dataset dimension}\label{subsec:fidelity_dataset}

\begin{figure*}[t]
    \centering
    \includegraphics[width=\textwidth]{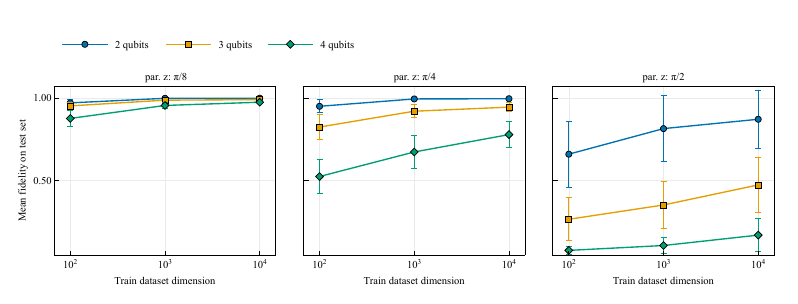}
    \caption{%
        Average fidelity of the LA-CPR method on the test set as a function
        of the training dataset size, for $z = \pi/8$ (left), $z = \pi/4$
        (center), and $z = \pi/2$ (right), with curves for $N_q = 2, 3, 4$
        qubits; error bars indicate one standard deviation of the fidelity
        across the unitaries of the test set.
        In all cases the fidelity increases monotonically with the dataset size
        and saturates beyond approximately $10^4$ elements.
        Larger values of $z$ make the reconstruction task progressively harder for the
        neural network, resulting in systematically lower fidelities.}
    \label{fig:fidelity_general}
\end{figure*}

Figure~\ref{fig:fidelity_general} shows the average fidelity the LA-CPR
method reaches on the test set, as a function of the training dataset size
on which it is trained, of the sampling parameter $z$, and of the number
of qubits $N_q$ in the system.
Each panel reports the fidelity curves, with the corresponding standard
deviations, as a function of the dataset size, one curve for each value of $N_q$.
In all configurations, the fidelity increases monotonically with the number of
training samples and tends to saturate beyond approximately $10^4$ elements,
with additional data yielding only marginal improvement.
This behavior is consistent with the well-known generalization properties of
neural networks, where performance improves with training data but eventually
reaches a regime of diminishing returns.

As expected, larger values of $z$ lead to a systematic degradation of the
fidelity.
A wider sampling interval $[-z, z]$ covers a larger and more diverse region of
the group manifold $SU(N)$, making the reconstruction task harder: the
neural network must learn to interpolate across a more heterogeneous set of
propagators, which reduces the precision of the generalization.
This effect is particularly evident when comparing the $z = \pi/8$ and
$z = \pi/2$ panels, where the latter shows a markedly lower fidelity across all
dataset sizes and qubit numbers.

A similar degradation is observed as the number of qubits $N_q$ increases.
At fixed $z$ and dataset size, the fidelity decreases monotonically when moving
from $N_q = 2$ to $N_q = 4$, as visible from the ordering of the curves in
every panel.
This trend is consistent with the exponential growth of the group dimension,
$\dim SU(2^{N_q}) = 4^{N_q} - 1$: realizing target unitaries on larger
registers requires control pulses with a richer time structure, and the
network must therefore interpolate the map $U \mapsto \Omega_{x,y}(t)$ over a
higher-dimensional and more intricate landscape, which is intrinsically harder
to learn from a finite training set.

These results indicate that, in practical applications of the LA-CPR method,
one should prefer sampling regions close to the identity of the Lie group,
i.e.\ small values of $z$, to obtain high-fidelity reconstructions without
requiring excessively large datasets.

\subsection{Comparison with standard gate decompositions}
To provide a meaningful benchmark for our Lie-algebra-based quantum optimal control (QOC)
reconstruction method, we evaluate the cost of implementing the same target unitaries through
standard gate decomposition as performed by a state-of-the-art quantum compiler. Specifically,
for each system size $N \in \{2, 3, 4\}$ qubits we draw $50$ target unitaries at
random from the set of propagators generated by the physical Hamiltonian of the system, and
decompose each of them into the native gate set $\{\mathrm{RZ}, \sqrt{X}, X, \mathrm{CNOT}\}$
using the Qiskit transpiler at optimization level 3. For each decomposed circuit we record the
following figures of merit: the total gate count, further broken down into the number of single-qubit gates
($n_{1q}$) and two-qubit gates ($n_{2q}$); the circuit depth, defined as the length of the
longest critical path through the circuit when gates acting on disjoint qubits are parallelized;
the estimated execution time $T$, defined as

\begin{figure*}[!t]
    \centering
    \includegraphics[width=\textwidth]{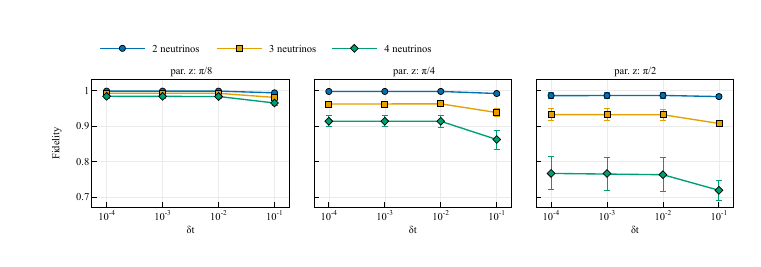}
    \caption{%
        Reconstruction fidelity of the LA-CPR method for the neutrino Trotter
        propagator $U_{\mathrm{neutr}} = e^{-i\,\delta t\, H_{\mathrm{neutr}}}$,
        shown as a function of the time step $\delta t$ and the number of
        neutrinos $N_n$, for $z = \pi/8$ (left), $z = \pi/4$ (center), and
        $z = \pi/2$ (right).
        Each marker is the mean fidelity obtained with $10$ independently
        trained models, and the error bar is the corresponding standard deviation.
        The fidelity is essentially independent of $\delta t$ up to
        $\delta t = 10^{-2}$ and decreases between $\delta t = 10^{-2}$ and
        $\delta t = 10^{-1}$; at every $\delta t$ it decreases with increasing $z$.}
    \label{fig:fidelity_neutrinos}
\end{figure*}

\begin{equation}
    T = n_{1q}\,\bar{t}_{1q} + n_{2q}\,\bar{t}_{2q},
\end{equation}
where $\bar{t}_{1q} = 10\,\mathrm{ns}$ and $\bar{t}_{2q} = 50\,\mathrm{ns}$ are the average
durations of single- and two-qubit gates, respectively; and the estimated circuit fidelity $F$,
defined as the product of the average per-gate fidelities over all gates in the circuit,
\begin{equation}
    F = \bar{F}_{1q}^{\,n_{1q}}\,\bar{F}_{2q}^{\,n_{2q}},
\end{equation}
with $\bar{F}_{1q} = 0.9996$ and $\bar{F}_{2q} = 0.9950$. These quantities, summarised in
Table~\ref{tab:gate_decomposition} as mean $\pm$ one standard deviation over the $50$ draws at
each $N_q$, serve as a reference against which the resource requirements of our method can be
compared. The last column reports, for the same system sizes, the fidelity of the LA-CPR
reconstruction at $z = \pi/4$.

\begin{table*}[t]
\centering
\caption{Gate decomposition of random unitaries for $N_q = 2, 3, 4$ qubits, compared with the
LA-CPR reconstruction fidelity at $z = \pi/4$. The decomposition columns report the mean $\pm$
one standard deviation over $50$ target unitaries drawn at random, for each $N_q$, from the
propagators generated by the physical Hamiltonian of the system. The LA-CPR column is likewise
reported as mean $\pm$ one standard deviation, over the unitaries of the test set, for a
training set of $10^4$ elements (the $z = \pi/4$ panel of Fig.~\ref{fig:fidelity_general}).}
\label{tab:gate_decomposition}
\begin{tabular*}{0.95\textwidth}{@{\extracolsep{\fill}}lccccccc@{}}
\toprule
$N_q$ &
\makecell{Total\\gates} &
\makecell{1q\\gates} &
\makecell{2q\\gates} &
Depth &
\makecell{Time\\{[ns]}} &
\makecell{Decomp.\\fidelity} &
\makecell{LA-CPR\\fidelity\\($z=\pi/4$)} \\
\midrule
2 & $17.0 \pm 0.7$  & $15.0 \pm 0.7$  & $2.0 \pm 0.0$  & $9.7 \pm 0.4$   & $250.2 \pm 7.3$    & $0.9841 \pm 0.0003$ & $0.996 \pm 0.003$ \\
3 & $90.7 \pm 4.2$  & $74.5 \pm 3.2$  & $16.2 \pm 1.0$ & $56.7 \pm 3.9$  & $1555.2 \pm 80.7$  & $0.8949 \pm 0.0055$ & $0.945 \pm 0.029$ \\
4 & $488.2 \pm 10.9$ & $402.4 \pm 9.3$ & $85.8 \pm 1.8$ & $298.2 \pm 6.7$ & $8312.0 \pm 178.0$ & $0.5539 \pm 0.0070$ & $0.779 \pm 0.079$ \\
\bottomrule
\end{tabular*}
\end{table*}

More broadly, this comparison underscores a general advantage of
control-based approaches over standard gate-decomposition pipelines, of which
LA-CPR is one particular instance.
By synthesizing complex unitaries directly as continuous control pulses, the
physical execution time scales far more favorably with $N_q$ than the rapidly
growing circuit depths produced by compilation into a discrete native gate set,
which in this benchmark inflate from ${\sim}250\,\mathrm{ns}$ at $N_q = 2$ to
${\sim}8300\,\mathrm{ns}$ at $N_q = 4$.
In the context of Trotterized quantum simulation, this allows each propagator
$e^{-i H \delta t}$ to be implemented as a single pulse rather than as a long
sequence of native gates, relaxing the pressure to slice $\delta t$ too finely
in order to keep the full evolution within the coherence window of present-day
qubits.

\section{Physical Application: Simulation of Neutrino Collective Oscillations}\label{subsec:neutrinos}

While the previous section assessed the LA-CPR method on random propagators, the
primary motivation of this work is the efficient reconstruction of control pulses
for propagators arising from the dynamics of physical quantum systems.
In general, the LA-CPR method cannot reconstruct arbitrary propagators, since the
sampling of the Lie algebra parameters is restricted to the interval
$[-z, z] \subsetneq [-\pi, +\pi]$, confining the training distribution to a
neighborhood of the group identity.
This limitation is, however, naturally compatible with Trotterized real-time
evolution, which is a standard technique in quantum simulation.
In the Trotter decomposition, the full time-evolution operator is factored into
short-time steps:
\begin{eqnarray}
U(T) \approx \prod_{t=0}^{T}e^{-iH_{\mathrm{syst}}(t)\,\delta t}=U_T U_{T-1}\cdots U_1 U_0,
\end{eqnarray}
where $T$ denotes the total evolution time.

In the limit $\delta t \rightarrow 0$, each factor $U_i \rightarrow \mathbbm{I}$;
consequently, by reducing $\delta t$, the individual Trotter propagators can be
confined to an arbitrarily small neighborhood of the identity in the Lie group.
The LA-CPR method, whose training set is constructed precisely from such a
neighborhood, is therefore directly applicable in this regime.

\begin{figure}[t]
    \centering
    \includegraphics[width=\columnwidth]{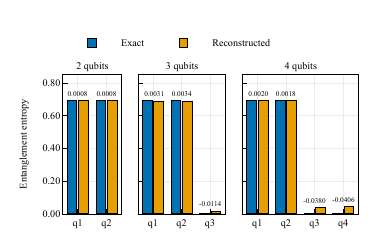}\\
    \includegraphics[width=\columnwidth]{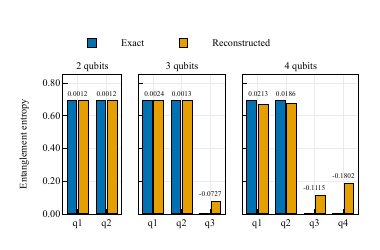}\\
    \includegraphics[width=\columnwidth]{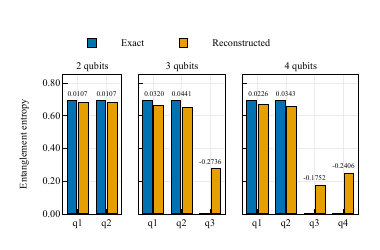}
    \caption{
        Single-qubit von Neumann entropy $S_k$ of the output state
        $\ket{\psi} = U\ket{\psi_0}$ for the exact (blue) and reconstructed
        (orange) neutrino propagators, shown for $N_n = 2, 3, 4$ neutrinos,
        $\delta t = 10^{-1}$,
        and $z = \pi/8$ (top), $z = \pi/4$ (middle), and $z = \pi/2$ (bottom).
        The number over each pair of bars is the difference
        $S_k^{\mathrm{exact}}-S_k^{\mathrm{rec}}$.
        The exact and reconstructed entropy profiles are in close agreement for
        small $z$, while the discrepancy grows with increasing $z$, mirroring
        the trend observed in the gate fidelity and confirming that
        reconstruction accuracy degrades as the training distribution broadens.}
    \label{fig:entanglement_neutrinos}
\end{figure}

\begin{figure}[t]
    \centering
    \includegraphics[width=\columnwidth]{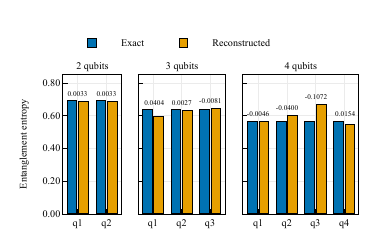}
    \caption{
        Single-qubit von Neumann entropy $S_k$ of the output state
        $\ket{\psi} = U\ket{W}$ for the exact (blue) and reconstructed
        (orange) neutrino propagators, shown for $N_n = 2, 3, 4$ neutrinos,
        $\delta t = 10^{-1}$, and $z = \pi/2$.
        Numbers over the bars as in Fig.~\ref{fig:entanglement_neutrinos}.}
    \label{fig:entanglement_W}
\end{figure}

As a concrete test case, we consider the collective flavor oscillations of
neutrinos arising from forward neutrino-neutrino scattering, following the
formulation of Refs.~\cite{amitrano2023neutrinos,Hall2021neutrinos}.
Neutrinos are approximated as two-flavor particles, with the electron flavor
$\nu_e$ and a single effective heavy flavor $\nu_x$ (representing a combination
of the $\mu$ and $\tau$ flavors).
Under this approximation, the flavor state of each neutrino,
$\ket{\psi_{\mathrm{neut}}}=\alpha\ket{\nu_e}+\beta\ket{\nu_x}$, maps naturally
onto a single-qubit state $\ket{\psi_q}=a\ket{0}+b\ket{1}$.
Each neutrino state therefore spans an $SU(2)$ space, and a system of $N_n$
neutrinos belongs to $SU(2^{N_n})$, which makes it a well-suited benchmark for
the LA-CPR method.
The flavor Hamiltonian decomposes into a one-body term $H_1$, accounting for
vacuum mixing, and a two-body neutrino-neutrino interaction term $H_2$,
generated by forward scattering:
\begin{eqnarray}
    H_{\mathrm{neutr}}&=&\underbrace{\sum_{i=1}^{N_n}\mathbf{b}\cdot \boldsymbol{\sigma}_i}_{H_1}+\underbrace{\sum_{i=1}^{N_n}\sum_{j=i+1}^{N_n}J_{ij}\,\boldsymbol{\sigma}_i \cdot \boldsymbol{\sigma}_j}_{H_2},
\end{eqnarray}
where bold symbols denote 3-dimensional vectors, $N_n$ is the number of
neutrinos, and $\boldsymbol{\sigma}_i = (\sigma_i^{(x)}, \sigma_i^{(y)},
\sigma_i^{(z)})$ collects the Pauli matrices acting on the $i$-th neutrino,
embedded in the full Hilbert space as
$\sigma_i^{(\alpha)}=\mathbbm{I}_1 \otimes \cdots \otimes \mathbbm{I}_{i-1}
\otimes \sigma^{(\alpha)} \otimes \mathbbm{I}_{i+1} \otimes \cdots \otimes
\mathbbm{I}_{N_n}$.
Following Ref.~\cite{amitrano2023neutrinos}, the vacuum mixing vector is
taken constant across $N_n$ and equal to
$\mathbf{b} = (0.38019,\,0,\,-0.92491)$.
The neutrino-neutrino couplings follow
$J_{ij} = 1 - \cos\theta_{ij}$ with geometric angles
$\theta_{ij} = \arccos(0.9)\,|i-j|/(N_n-1)$,
so $J_{ij}$ depends only on the pair distance $|i-j|$ and on $N_n$.
The resulting values are summarised in
Table~\ref{tab:neutrino_params}.

\begin{table}[h]
    \centering
    \caption{Neutrino-neutrino couplings $J_{ij}$ entering
        $H_{\mathrm{neutr}}$, listed by pair distance $|i-j|$
        for $N_n = 1, 2, 3, 4$.}
    \label{tab:neutrino_params}
    \begin{tabular*}{0.95\columnwidth}{@{\extracolsep{\fill}}cccc@{}}
    \toprule
    $N_n$ & $|i-j|$ & $\theta_{ij}$ [rad] & $J_{ij}$ \\
    \midrule
    1 & --  & --    & --    \\
    2 & 1    & 0.4510 & 0.1000 \\
    3 & 1    & 0.2255 & 0.0253 \\
    3 & 2    & 0.4510 & 0.1000 \\
    4 & 1    & 0.1503 & 0.0113 \\
    4 & 2    & 0.3007 & 0.0449 \\
    4 & 3    & 0.4510 & 0.1000 \\
    \bottomrule
    \end{tabular*}
\end{table}

We benchmark the LA-CPR method on the single-step Trotter propagator:
\begin{eqnarray}
    U_{\mathrm{neutr}}=e^{-i\,\delta t\, H_{\mathrm{neutr}}},
\end{eqnarray}
for $N_n = 2, 3, 4$ neutrinos, whose propagators belong to $SU(4)$, $SU(8)$,
and $SU(16)$, respectively.
For each system size, the time step is varied over
$\delta t \in \{10^{-4}, 10^{-3}, 10^{-2}, 10^{-1}\}$, so as to probe identity
neighborhoods of progressively increasing width.
The results are presented in Fig.~\ref{fig:fidelity_neutrinos}, where each panel
corresponds to a fixed value of $z$ and shows the reconstruction
fidelity as a function of $\delta t$ with a line for every number of neutrinos $N_n$.
Since the target propagator is fixed for each $N_n$ and $\delta t$, the spread
of the fidelity cannot be estimated over a set of test unitaries, as in
Sec.~\ref{subsec:fidelity_dataset}.
Instead, the LA-CPR networks are trained $10$ times independently on the same
dataset of $10\,000$ elements, and the same propagator is reconstructed with
each of the $10$ trained models; each point in
Fig.~\ref{fig:fidelity_neutrinos} is the mean fidelity, and the error bar its
standard deviation.
The fidelity is essentially independent of $\delta t$ for $\delta t \le 10^{-2}$
and decreases between $\delta t = 10^{-2}$ and $\delta t = 10^{-1}$, more
markedly for larger $N_n$.
At every $\delta t$, the fidelity decreases with increasing $z$, consistently
with the results of Sec.~\ref{subsec:fidelity_dataset}: larger values of $z$ broaden the
training distribution, increasing the difficulty of the reconstruction task.

To further validate the reconstruction at the state level, we assess whether the
reconstructed controls faithfully reproduce the entanglement structure generated
by the exact neutrino propagators.
We prepare the input state
$|\psi_0\rangle = (|10\cdots0\rangle + |01\cdots0\rangle)/\sqrt{2}$,
a single-excitation superposition across the first two qubits, and evolve it
under both the exact and reconstructed unitary operators.
For each output state $|\psi\rangle = U|\psi_0\rangle$, we compute the
single-qubit von Neumann entropy
$S_k = -\mathrm{Tr}[\rho_k \ln \rho_k]$, where
$\rho_k = \mathrm{Tr}_{\bar{k}}\big(|\psi\rangle\langle\psi|\big)$ is the reduced
density matrix of qubit $k$, with $\bar{k}$ denoting the set of all qubits other
than $k$.
The results are shown in Fig.~\ref{fig:entanglement_neutrinos} for
$N_n = 2, 3, 4$ neutrinos and the three values of $z$ considered throughout
this work.
For $z = \pi/8$, the exact and reconstructed entropy profiles agree closely
across all system sizes, confirming that the reconstructed controls faithfully
capture the entanglement dynamics of the neutrino propagators.
As $z$ increases, the agreement progressively deteriorates, consistent with the
decrease in gate fidelity observed in Fig.~\ref{fig:fidelity_neutrinos}: a
broader training distribution reduces reconstruction accuracy and, consequently,
the ability of the reconstructed controls to reproduce the correct entanglement
structure.

Fig.~\ref{fig:entanglement_neutrinos} also shows that the largest discrepancies
occur on qubits 3 and 4.
The size of the discrepancy on these qubits is likely set by the behaviour of
the entropy near a pure state, rather than by a qubit-specific failure of the
reconstruction.
Indeed, the input state entangles only the first two qubits, so that $\rho_1$
and $\rho_2$ are close to the maximally mixed state, while the remaining qubits
start in $\ket{0}$ and are only weakly entangled by the exact evolution, so that
their reduced states remain close to pure.

For a single qubit, $S_k$ depends only on the smaller eigenvalue $p$ of
$\rho_k$, through $S(p) = -p\ln p-(1-p)\ln(1-p)$.
According to this expression, at $p=1/2$ the entropy is stationary, and a
deviation $\delta$ lowers it only by $2\delta^2$.
As $p\to0$ the slope $\mathrm{d}S/\mathrm{d}p=\ln[(1-p)/p]$ diverges and
$S\simeq p\ln(1/p)$, so a small population transfer onto a nearly pure qubit
produces a large entropy increase.

This interpretation is supported by repeating the analysis with the $W$ state
$\ket{W}=(\ket{10\cdots0}+\ket{01\cdots0}+\cdots+\ket{0\cdots01})/\sqrt{N_n}$,
for which every reduced state has $p=1/N_n$, far from $p=0$.
The results are shown in Fig.~\ref{fig:entanglement_W}, for simplicity only for
$z=\pi/2$.
The differences are spread over all qubits, and the largest ($0.107$, for
$N_n=4$) is smaller than those found on the unentangled qubits for the
two-qubit superposition ($0.175$--$0.274$).
This state-level analysis provides an independent validation of the
reconstruction quality, complementary to the average gate fidelity $\bar{F}$.
It is also worth emphasizing that the neural networks were trained exclusively on
randomly sampled propagators and were never exposed to the neutrino propagators
during training.

These results suggest that the LA-CPR method can effectively reconstruct the
control pulses for unitaries in $SU(N)$ on $N_q$ qubits, at least within
the regime of sufficiently small Trotter steps, a condition naturally met in
Trotterized quantum simulation.

\section{Conclusions}

We have introduced the Lie Algebra-based Control Pulse Reconstruction (LA-CPR)
method, a hybrid approach that combines the mathematical structure of Lie groups
with the generalization capabilities of feed-forward neural networks to address
one of the central bottlenecks in quantum optimal control: the need to
independently re-optimize control pulses for every new target propagator.

The key insight underlying LA-CPR is that all propagators of an $N_q$-qubit
system belong to $SU(N)$, a compact Lie group that can be systematically
sampled via its associated Lie algebra.
By exploiting the Pauli tensor basis decomposition of $\mathfrak{su}(N)$
and restricting the family of \emph{synthesized} unitaries to those generated
by a linear chain topology with Ising $ZZ$ coupling -- a structural choice
that reduces the number of sampling parameters while preserving universality,
and which is independent of the underlying qubit Hamiltonian -- we obtain a
compact parametrization that still covers the full group through iterated
Lie brackets.
A dataset of GRAPE-optimized control pulses is computed once for a representative
sample of group elements drawn from a neighborhood of the identity, and two
feed-forward neural networks are trained to map each propagator to its
corresponding $\Omega_x(t)$ and $\Omega_y(t)$ control envelopes.

The numerical experiments on systems of $N_q = 2,3,4$ qubits demonstrate that
LA-CPR achieves high reconstruction fidelity with training datasets of
approximately $10^4$ elements, beyond which additional samples yield only
marginal improvement.
As expected, restricting the sampling interval $[-z, z]$ to smaller values of
$z$ -- corresponding to a tighter neighborhood of the group identity --
systematically improves reconstruction quality, since the neural network is
required to interpolate over a more homogeneous region of the group manifold.

A comparison with standard gate-decomposition pipelines further highlights a
structural advantage of control-based methods, of which LA-CPR is one instance:
synthesizing each target unitary as a single continuous pulse achieves higher
fidelities in much shorter physical times, relaxing the need for very small
Trotter steps to stay within the coherence window of present-day qubits.

The application to neutrino collective oscillations provides a compelling proof
of concept for the use of LA-CPR in quantum simulation.
Despite the fact that the neural networks were trained exclusively on
hardware-specific random propagators, the method successfully reconstructed
control pulses for the physical Trotter propagators of $N_n = 2,3,4$ neutrino
systems across a range of time steps, without any retraining or system-specific
adaptation.
This result confirms the key practical advantage of LA-CPR: a single offline
training phase, tied to the hardware Hamiltonian rather than to any particular
target system, produces a reusable model that can be deployed for any quantum
simulation task on the same device.
This makes LA-CPR particularly well suited to the simulation of many-body
systems, where large ensembles of distinct propagators must be processed
efficiently.

All fidelities reported in this work refer to closed-system dynamics and
quantify the reconstruction error alone.
Once generated, an LA-CPR pulse is executed on the device like any other
optimal-control pulse: decoherence acts during a window of fixed duration
$\tau$, identical for every target unitary, and fluctuations of the drive act
directly on $\Omega_x(t)$ and $\Omega_y(t)$.
Robustness can therefore be built into the method at the data-generation stage.
Because LA-CPR does not depend on the optimizer that produces the training
pulses, open-system and ensemble-robust variants of
GRAPE~\cite{chen2025robust_open}, or optimal-control schemes that include drive
noise in the optimization~\cite{aroch2024controller_noise}, can replace the
closed-system optimization used here, with no change to the sampling of the
group or to the network architecture.
A drift of the hardware parameters changes the map learned by the network; it
can be handled by periodic retraining, or by including the drifting parameters
among the network inputs.
A quantitative assessment of the reconstruction fidelity under realistic noise
models, and of whether the robustness of the training pulses carries over to
the interpolated ones, is left to future work.

Looking ahead, several natural extensions of the method merit investigation.
Scaling to larger qubit registers will require addressing the exponential growth
of the input dimension and the increased diversity of the group manifold;
techniques such as structured input representations, locality-aware network
architectures, or hierarchical decompositions may prove beneficial in this
regime.
It would also be interesting to explore whether training on correlated,
physically motivated datasets -- rather than uniformly random samples -- could
improve fidelity for specific classes of target systems.
Finally, a systematic study of the interplay between Trotter step size, sampling
radius $z$, and reconstruction fidelity would help establish practical guidelines
for deploying LA-CPR within larger quantum simulation workflows.

\newpage

\appendix

\section{\texorpdfstring{Pauli Basis Decomposition: $N_q=3$}{Pauli Basis Decomposition: Nq=3}}\label{App:pauli_basis_Nq3}

For $N_q=3$, the algebra $\mathfrak{su}(8)$ has dimension
$4^3 - 1 = 63$.
The Hamiltonian expands as
\begin{align}
    H &= \underbrace{\sum_{i=1}^3 \sum_{a}
         \gamma_a^{(i)} \sigma_a^{(i)}}_{\text{1-qubits: 9 terms}}
       + \underbrace{\sum_{i<j} \sum_{a,b}
         \gamma_{ab}^{(ij)}\, \sigma_a^{(i)} \otimes \sigma_b^{(j)}}_{\text{2-qubits: 27 terms}}
    \nonumber\\
      &\quad + \underbrace{\sum_{a,b,c}
         \theta_{abc}\,
         \sigma_a^{(1)} \otimes \sigma_b^{(2)} \otimes \sigma_c^{(3)}}_{\text{3-qubits: 27 terms}}.
\end{align}

\begin{table}[h]
\centering
\caption{Pauli basis for $N_q=3$, $\dim\mathfrak{su}(8)=63$.}
\begin{tabular*}{0.95\columnwidth}{@{\extracolsep{\fill}}lll@{}}
\toprule
Locality & Operators & No.\ of terms \\
\midrule
1-qubits & $\sigma_a^{(i)}$, $i\in\{1,2,3\}$ & 9 \\
2-qubits & $\sigma_a^{(i)}\otimes\sigma_b^{(j)}$, $i<j$ & 27 \\
3-qubits & $\sigma_a^{(1)}\otimes\sigma_b^{(2)}\otimes\sigma_c^{(3)}$ & 27 \\
\midrule
\textbf{Total} & & \textbf{63} \\
\bottomrule
\end{tabular*}
\end{table}

\section{Simplification of Lie Algebra}\label{App:simpl_lie_alg}

\subsection{Chain topology simplification}
We show here, for $N_q = 3$, how nested commutators of elements of the
space $\mathcal{V}$ of Eq.~\eqref{eq:dynamic_algebra_g} generate the
directions of $\mathfrak{su}(8)$ missing from $\mathcal{V}$, i.e.\ the
Pauli strings acting non-trivially on both qubits 1 and 3.
A single commutator does not suffice: for Pauli strings in
$\mathfrak{su}(4)_{12}$ and $\mathfrak{su}(4)_{23}$, the factors on
qubits 1 and 3 always commute, so the bracket is non-zero only if the
factors on qubit 2 anticommute, and their product is then a third Pauli
matrix, never the identity. Two nested brackets suffice. Consider the
generators
\begin{align}
    A &= i\,(\sigma_x \otimes \sigma_y \otimes \mathbbm{I}) \in \mathfrak{su}(4)_{12}, \notag \\
    B &= i\,(\mathbbm{I} \otimes \sigma_z \otimes \sigma_z) \in \mathfrak{su}(4)_{23},
\end{align}
both elements of $\mathcal{V}$. Their commutator is
\begin{align}
    [A, B] &= -\,[\sigma_x \otimes \sigma_y \otimes \mathbbm{I},\;
                 \mathbbm{I} \otimes \sigma_z \otimes \sigma_z] \notag \\
           &= -\,\sigma_x \otimes (\sigma_y\sigma_z - \sigma_z\sigma_y) \otimes \sigma_z \notag \\
           &= -\,\sigma_x \otimes 2i\sigma_x \otimes \sigma_z \notag \\
           &= -2i\,(\sigma_x \otimes \sigma_x \otimes \sigma_z),
\end{align}
a three-body Pauli string outside $\mathcal{V}$. With
$C = i\,(\sigma_y \otimes \sigma_x \otimes \mathbbm{I}) \in \mathfrak{su}(4)_{12}$,
\begin{align}
    [C, [A, B]] &= 2\,[\sigma_y \otimes \sigma_x \otimes \mathbbm{I},\;
                      \sigma_x \otimes \sigma_x \otimes \sigma_z] \notag \\
                &= 2\,(\sigma_y\sigma_x - \sigma_x\sigma_y) \otimes \sigma_x^2 \otimes \sigma_z \notag \\
                &= 2\,(-2i\sigma_z) \otimes \mathbbm{I} \otimes \sigma_z \notag \\
                &= -4i\,(\sigma_z \otimes \mathbbm{I} \otimes \sigma_z).
    \label{eq:nested_comm}
\end{align}
The operator $\sigma_z \otimes \mathbbm{I} \otimes \sigma_z$ acts non-trivially on qubits 1 and 3
simultaneously, while being the identity on the mediating qubit 2.
This is a \emph{long-range} Pauli string---one that has no support in
either $\mathfrak{su}(4)_{12}$ or $\mathfrak{su}(4)_{23}$ alone---and
it lies outside $\mathcal{V}$. Repeating
this procedure for all inequivalent pairs of nearest-neighbor Pauli strings,
one systematically generates all $4^3 - 1 = 63$ independent traceless
Pauli strings on three qubits.

\subsection{ZZ coupling simplification}
We show here how the restricted space
\begin{equation}
    \mathcal{V}_{\mathrm{ZZ}}
    = \mathrm{span}_{\mathbb{R}}\bigl\{\,
        i\sigma_a^{(i)},\;
        i\sigma_z^{(i)}\otimes\sigma_z^{(i+1)}
    \bigr\}
\end{equation}
still closes to $\mathfrak{su}(N)$ under iterated Lie brackets,
even though only one of the nine two-body generators of
$\mathfrak{su}(4)_{i,i+1}$ (the other six are single-qubit) is
retained natively. The missing interaction types are recovered by commuting the $ZZ$ generator
with single-qubit rotations. For instance, the $YZ$-type term is
obtained as
\begin{align}\label{eq:YZ_commutator}
    &\bigl[i\,\sigma_z^{(i)}\otimes\sigma_z^{(i+1)},\;
          i\,\sigma_x^{(i)}\otimes\mathbbm{I}^{(i+1)}\bigr] \nonumber\\
    &\quad = -\bigl[\sigma_z^{(i)},\sigma_x^{(i)}\bigr]
        \otimes\sigma_z^{(i+1)} \nonumber\\
    &\quad = -2i\,\sigma_y^{(i)}\otimes\sigma_z^{(i+1)},
\end{align}
where we used $[\sigma_z, \sigma_x] = 2i\sigma_y$ and the
mixed-product property of the tensor commutator. Analogous
commutators with $\sigma_y^{(i)}$ or with operators on site $i{+}1$
generate the remaining $XZ$, $ZX$, $ZY$ Pauli strings, and one
further bracket with a single-qubit generator turns these into the
$XX$, $XY$, $YX$, $YY$ couplings, e.g.
$\bigl[i\,\mathbbm{I}^{(i)}\otimes\sigma_y^{(i+1)},\;
i\,\sigma_x^{(i)}\otimes\sigma_z^{(i+1)}\bigr]
= -2i\,\sigma_x^{(i)}\otimes\sigma_x^{(i+1)}$.
Combined with the chain-topology construction of the previous subsection, this recovers all $4^{N_q} - 1$ traceless
Pauli strings, establishing that $\mathcal{V}_{\mathrm{ZZ}}$
generates $\mathfrak{su}(N)$.

\section{LA-CPR Method technical details}\label{App:LA-CPR_details}

\subsubsection{LA-CPR Preparation}\label{sec:LA_CPR_preparation}
LA-CPR method requires a preparation phase before its actual use.
Let's consider the $SU(N)$ group with its associated Lie algebra
$\mathfrak{su}(N)$. We should:

\begin{enumerate}\label{App.sec:LA-CPR prepatation}
    \item Provide the experimental parameters of the device in use, i.e.\ the
        specific qubits Hamiltonian of Eq.~\eqref{eq:H_qubit_tot}.
    \item Define the basis set $\{T_p\}$ of $\mathfrak{su}(N)$ as described in
        Sec.~\ref{subsec:lie_theory}.
    \item Compute the set of matrices $\{U_j\}$ via Eq.~\eqref{eq:LA_sampling_U}
        sampling the parameters $\gamma_l$ from a uniform distribution in the
        interval $[-z,z]$ with $z \in (0,\pi]$.
    \item Compute the controls $\epsilon_j^k(t)$, using GRAPE algorithm, to solve
        the formal minimization problem of Eq.~\eqref{eq:general_opt_problem} for
        each $U_j$ in the dataset. We thus obtain $2$ datasets of controls
        representing $\Omega_x(t)$ and $\Omega_y(t)$ of the qubits systems
        Eq.~\eqref{eq:H_ctrl}.
    \item Prepare the inputs and the outputs for the neural network.
        Since the neural network can only take vectors as inputs, each matrix
        $U_j$ is flattened stacking the real and the imaginary part of the matrix
        in a single vector, i.e.\ $U_j^{flat}= [\Re U_j^{1,1},\ldots\Re
        U_j^{N,N},\Im U_j^{1,1},\ldots\Im U_j^{N,N}]$.
        The outputs instead are the controls themselves, which are already in the
        shape of single vectors.
    \item Define $2$ feed-forward neural networks (FFNN), one for each output
        dataset, with an appropriate architecture.
        Train each FFNN to link $U_j^{flat}$ to the corresponding control
        $\epsilon^k_j(t)$, i.e.\ $\epsilon^k_j=f_k(U_j^{flat})$.
    \end{enumerate}

\subsubsection{LA-CPR Use}
After the preparation phase, the use of the LA-CPR method is straightforward.
\begin{enumerate}
    \item Take any arbitrary matrix $\tilde{U} \in SU(N)$ not in the training
        dataset.
    \item Flatten $\tilde{U}$ in a vector, $\tilde{U} \rightarrow \tilde{U}^{flat}$.
    \item Insert $\tilde{U}^{flat}$ in the $2$ neural networks to obtain the
        corresponding controls $\tilde{\Omega_a}=f_a(\tilde{U}^{flat})$
        for $a=x,y$.
\end{enumerate}
The matrix $\tilde{U}$ can be random or come from any quantum system we are
interested in.
The LA-CPR method is able to provide the control pulses for that specific
propagator without performing an optimization and without requiring any knowledge
of the quantum system itself.

\subsubsection{Neural Networks}

The neural networks used in this work are feedforward neural networks.
The input, as specified in Step 5 of Sec.~\ref{App.sec:LA-CPR prepatation},
is the flattened representation $U_j^{\text{flat}}$ of a propagator $U_j$ in
matrix form, and the output is the control pulse $\Omega_a(t)$ associated with
that propagator.
The network architecture consists of two hidden layers of 500 neurons each, with
ReLU activation functions for the hidden layers.
After each hidden layer, a Dropout layer with a rate of $0.1$ is applied to
reduce overfitting.

No normalization is applied to either the input or output data.
The entries of the flattened propagator $U_j^{\text{flat}}$ are naturally bounded
in $[-1, 1]$, since the real and imaginary parts of a unitary matrix satisfy this
constraint.
Similarly, the control amplitudes $\Omega_a(t)$ oscillate within a comparable
range, making explicit normalization unnecessary.
Should different operating conditions require it, a normalization step could
straightforwardly be introduced as a preprocessing stage.

Training is performed by minimizing a mean squared error (MSE) loss function
using the Adam optimizer.
The maximum number of epochs is set to $100$, with an early stopping callback
that halts training if the validation loss does not decrease for $6$ consecutive
epochs.

The choice of the network architecture -- namely the number of hidden
layers and neurons per layer -- was determined through a grid search
over a range of candidate configurations.
The best-performing architecture was selected based on the validation MSE, and
the results of this search are reported in Fig.~\ref{fig:heatmap_architectures}.

\section{Neural network architecture selection}\label{App.nn_architect}

\begin{figure}[h]
    \centering
    \begin{subfigure}[t]{\columnwidth}
        \centering
        \includegraphics[width=\linewidth]{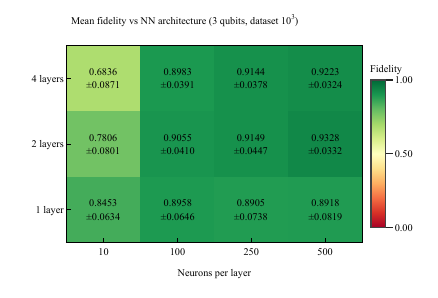}
        \caption{Reconstruction fidelity.}
        \label{fig:heatmap_fidelity}
    \end{subfigure}
    \begin{subfigure}[t]{\columnwidth}
        \centering
        \includegraphics[width=\linewidth]{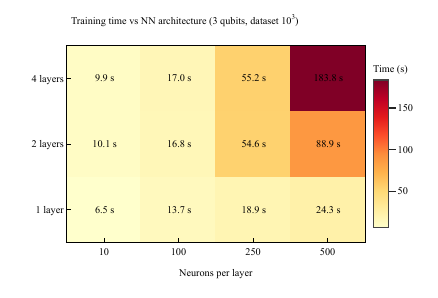}
        \caption{Training time.}
        \label{fig:heatmap_training_time}
    \end{subfigure}
    \caption{%
        Grid search over neural network architectures: reconstruction fidelity
        (top) and training time (bottom) for all combinations of
        $N_l \in \{1,2,4\}$ hidden layers and layer width
        $N_D \in \{10,100,250,500\}$, trained on $10\,000$ samples with
        $z=\pi/4$ for 3 qubits and evaluated on a 200-element test set.}
    \label{fig:heatmap_architectures}
\end{figure}

Since there is no general theoretical recipe for choosing the best neural
network architecture, the selection is carried out empirically through a
grid search over depth and width.
Networks with $N_l \in \{1,2,4\}$ hidden layers and $N_D \in \{10,100,250,500\}$
neurons per layer are trained on a dataset of $10\,000$ elements with
$z=\pi/4$ for 3 qubits, and evaluated on a test set of 200 elements.
The results are shown in Fig.~\ref{fig:heatmap_architectures}: the left panel
reports the average reconstruction fidelity (with standard error) for each
$(N_l, N_D)$ combination, while the right panel shows the corresponding
training time.

The two panels together allow a joint assessment of accuracy and computational
cost, facilitating the identification of an efficient architecture.
As expected, increasing both the number of layers $N_l$ and the layer width
$N_D$ generally improves the reconstruction fidelity.
However, this gain is not linear: the marginal improvement in fidelity diminishes
as the network grows larger, while the training time increases sharply.
At the low end, networks with only $N_D = 10$ neurons per layer consistently
underperform regardless of depth, indicating that this width is insufficient to
capture the complexity of the mapping between unitary targets and control pulses.
On the other hand, the largest configurations do not systematically outperform
intermediate ones and can even degrade slightly, likely due to training
difficulties and overfitting on the finite dataset.

Based on this analysis, the architecture with $N_l = 2$ hidden layers and
$N_D = 250$ neurons per layer is selected and used for all experiments
in this work.
This configuration achieves a favorable trade-off between fidelity
(comparable to heavier architectures) and training cost (lower than heavier architectures).
It should be noted that this choice is not universal and can be adjusted in
each specific implementation depending on the available computational resources
and the desired accuracy.

\FloatBarrier
\bibliography{bibliography}

\end{document}